\documentclass[aps,prd,groupedaddress,preprintnumbers,nofootinbib,twocolumn,superscriptaddress]{revtex4-1}

\usepackage{amsmath, amssymb}
\usepackage{mathtools}
\usepackage[pdftex]{graphicx}
\usepackage[colorlinks]{hyperref}
\usepackage{color}
\usepackage{setspace, enumitem}
\usepackage[normalem]{ulem}
\usepackage{enumitem}
\usepackage{physics}
\usepackage[dvipsnames]{xcolor}
\usepackage{simpler-wick}
\usepackage{float}
\usepackage{amsmath}
\usepackage{hyperref}
\usepackage[capitalise]{cleveref}
\usepackage{youngtab}
\usepackage{mathdots}
\usepackage{dsfont}
\usepackage[graphicx]{realboxes}
\usepackage{adjustbox}
\usepackage{rotating}
\usepackage{mathdots}
\usepackage{cancel}

\newcommand{\fstop}{\, .}
\renewcommand{\Tr}{\text{Tr}}

\newcommand{\Nc}{N_C}
\newcommand{\Nf}{N_F}

\newcommand{\Qb}{\widetilde{Q}}
\newcommand{\Bn}{B_n}

\newcommand{\Bo}{B_0}
\newcommand{\Bi}{B_1}
\newcommand{\Bii}{B_2}

\newcommand{\Sp}{\text{Sp}}
\newcommand{\SU}{\text{SU}}
\newcommand{\U}{\text{U}}
\newcommand{\diag}{\text{diag}}

\definecolor{HMG}{RGB}{212,175,55} 
\definecolor{BSB}{RGB}{4, 55, 242} 
\definecolor{JMLG}{rgb}{0.172549, 0.627451, 0.172549} 
\definecolor{JWG}{RGB}{153, 50, 204} 
\definecolor{GSG}{RGB}{243, 112, 243} 

\begin{document}
\title{Exact Results in Softly-Broken Supersymmetric Chiral Gauge Theories with Flavor}

\author{Jacob M. Leedom}
\email{leedom@fzu.cz}
\affiliation{CEICO, Institute of Physics of the Czech Academy of Sciences\\
Na Slovance 2, 182 00 Prague 8, Czech Republic}

\author{Hitoshi Murayama}
\email{hitoshi@berkeley.edu} 
\thanks{Hamamatsu Professor}
\affiliation{Theory Group, Lawrence Berkeley National Laboratory, Berkeley, CA 94720, USA}
\affiliation{Berkeley Center for Theoretical Physics, University of California, Berkeley, CA 94720, USA}
\affiliation{Kavli Institute for the Physics and Mathematics of the Universe (WPI),
University of Tokyo, Kashiwa 277-8583, Japan}

\author{Gup Singh}
\email{gupsingh@berkeley.edu}
\affiliation{Theory Group, Lawrence Berkeley National Laboratory, Berkeley, CA 94720, USA}
\affiliation{Berkeley Center for Theoretical Physics, University of California, Berkeley, CA 94720, USA}

\author{Bethany Suter}
\email{bethany\_suter@berkeley.edu}
\affiliation{Theory Group, Lawrence Berkeley National Laboratory, Berkeley, CA 94720, USA}
\affiliation{Berkeley Center for Theoretical Physics, University of California, Berkeley, CA 94720, USA}

\author{Jason Wong}
\email{jtwong71@berkeley.edu}
\affiliation{Theory Group, Lawrence Berkeley National Laboratory, Berkeley, CA 94720, USA}
\affiliation{Berkeley Center for Theoretical Physics, University of California, Berkeley, CA 94720, USA}

\begin{abstract}
    We present exact results in softly-broken supersymmetric $\SU(\Nc)$ chiral gauge theories with charged fermions in one antisymmetric,  $N_F$ fundamental, and $\Nc+\Nf-4$ anti-fundamental representations. We achieve this by considering the supersymmetric version of these theories and utilizing anomaly mediated supersymmetry breaking at a scale $m\ll\Lambda$ to generate a vacuum. The connection to non-supersymmetric theories is then conjectured in the limit $m\rightarrow\infty$.  For odd $\Nc$, we determine the massless fermions and unbroken global symmetries in the infrared. For even $\Nc$, we find global symmetries are non-anomalous and no massless fermions. In all cases, the symmetry breaking patterns differ from what the tumbling hypothesis would suggest. 
\end{abstract}
    
\maketitle

\section{Introduction}
\label{sec:intro}
Gauge theories form the theoretical foundation of fundamental interactions, but numerous questions on their characteristics persist. This is particularly true in the strongly-coupled regime, where calculations are scarce due to a lack of proper tools. In addition, even lattice gauge theories, arguably the best tool to study the strongly-coupled regime, cannot be used for chiral gauge theories still to this day. Another tool is the tumbling hypothesis \cite{Georgi:1979md,Raby:1979my,Goity:1985tf}, which assumes fermion bilinear condensate in the most attractive channel and can predict specific symmetry breaking patterns, but it had not been tested in any way.

One fruitful technique has been to consider gauge theories with supersymmetry (SUSY), which enjoy greater control due to a union of holomorphicity and global symmetries. One could hope that this success could be translated to non-SUSY theories via the introduction of SUSY-breaking deformations that nonetheless maintain some level of calculability. However, many such approaches either surrender control or do not fully map to their non-SUSY analogues. An exception is \textit{anomaly-mediated supersymmetry breaking}~(AMSB) \cite{Randall:1998uk,Giudice:1998xp}, which escapes the above pitfalls thanks to  \textit{ultraviolet insensitivity}~\cite{Pomarol:1999ie,Boyda:2001nh}. Namely, the SUSY breaking effects are controlled at all energy scales.

When this technique was applied to chiral gauge theories, the exact results did not support the tumbling hypothesis \cite{Csaki:2021xhi,Csaki:2021aqv,Kondo:2022lvu}. It would be helpful to draw general lessons by studying more examples. The goal this work is to study examples where tumbling predicts specific symmetry breaking patterns. As shown below, this technique again finds exact results whose symmetry breaking patterns do not agree with tumbling.

The strategy of AMSB is to deform a $\mathcal{N}=1$ SUSY gauge theory with matter supermultiplets $\{\phi_I\}$ via the 
inclusion of the Weyl compensator superfield 
\begin{equation}
    \Phi = 1+\theta^2 m,
\end{equation}
where $\theta$ is the typical Grassmanian variable for $\mathcal{N}=1$ superspace. Masses of order $m$ are thus introduced for the gauginos and squarks\footnote{Here squarks is used as a catch-all term for the scalar components of the supermultiplets $\{\phi_I\}$.}. More concretely, the typical SUSY Lagrangian is altered to include the Weyl compensator field in the following manner:
\begin{equation}
    \mathcal{L} \supset \int d^4\theta \Phi^*\Phi K + \int d^2\theta
    \Phi^3W + \int d^2\theta^*\Phi^{*3}W^*,
\label{eq:AMSBsuperspace}
\end{equation}
where $K$ and $W$ are the K\"{a}hler potential and superpotential, respectively. The scalar potential is then given by a sum of three terms:
\begin{equation}
    \begin{aligned}
    V &= V_F + V_D + V_{AMSB},\\
    V_{AMSB} &= m\bigg(\phi_I\frac{\partial W}{\partial\phi_I}-3W \bigg) + \text{ h.c.}\\
\end{aligned}
\label{eq:AMSBpot}
\end{equation}
Here $V_F$ and $V_D$ are the usual F-term and D-term scalar potentials, respectively, determined via the inverse K\"{a}hler metric $K_{I\bar{J}}$
and gauge kinetic function. The remaining term, $V_{AMSB}$, arises from the inclusion of the AMSB terms in~\cref{eq:AMSBsuperspace}. The D-term potential in~\cref{eq:AMSBpot} vanishes so long as the D-flat condition
\begin{equation}
2AA^\dag + Q Q^\dag - \Qb^* \widetilde{Q}^T = \sigma \mathds{1}
\label{eq:DFlat}
\end{equation}
is satisfied. A non-supersymmetric vacuum can then appear via a balancing of the remaining two terms and the squarks obtain non-zero vacuum expectation values (vevs). 

Thus far, we have only made statements about SUSY theories with a particular deformation. To connect with bonafide non-supersymmetric chiral gauge theories, we must consider the limit $m\rightarrow \infty$ so that the squarks and gauginos become decoupled. Assuming no phase transition occurs as $m$ grows larger than $\Lambda$, the AMSB-deformed theory and the non-SUSY theory will lie in the same universality class. If this holds, then features of the AMSB-deformed theories such as the existence of global symmetries and massless fermions will carry over to the non-SUSY theory. The above technique was first utilized for QCD-like theories in~\cite{Murayama:2021xfj}, but has since been extended to several classes of gauge theories~\cite{Csaki:2021xhi,Csaki:2021aqv,Kondo:2022lvu,Murayama:2021rak,Csaki:2021xuc,Csaki:2021xuc,Csaki:2022cyg,Goh:2025}.

In this paper, we apply the above approach to chiral gauge theories with extended flavor symmetries. In particular, we obtain exact results in softly-broken supersymmetric chiral gauge theories determined by a pair of integers $\{\Nc,\Nf\}$ such that the gauge group is $\SU(\Nc)$ and the charged matter content consists of i) $\Nf$ fermions in the fundamental representation, ii) $(\Nc+\Nf-4)$ fermions in the anti-fundamental representation, and iii) a single fermion in the anti-symmetric representation. This matter content ensures the cancellation of local gauge anomalies. 

The $\mathcal{N}=1$ supersymmetric version of these theories were discussed in~\cite{Poppitz:1995fh,Pouliot:1995me}. In the ultraviolet (UV), their matter sectors consist of $\Nf$ chiral supermultiplets $Q_\alpha^a$ in the fundamental representation, $(\Nc+\Nf-4)$ chiral multiplets $\widetilde{Q}^\alpha_i$ in the anti-fundamental representation, and a chiral multiplet $A_{\alpha\beta}$ in the anti-symmetric tensor representation. We use Greek letters for gauge indices ($\alpha,\beta,\gamma \in \{1,..,\Nc\}$) and Latin letters for flavor indices ($a,b\in \{1,..,\Nf\}$ and $i,j\in \{1,..,\Nc+\Nf-4\}$).  

The above field content implies that the theories have a global symmetry group
\begin{align}
    \lefteqn{\mathcal{G}_{UV}(\Nc,\Nf)} \nonumber \\ 
    &:=\SU(\Nf)\times \SU(\Nc+\Nf-4) \times \U(1)_1\times \U(1)_2.
\label{eq:UVglobal}
\end{align}
The charges of the fields are given in~\cref{tab:charges}. There is also a global R-symmetry $\U(1)_R$, but this is broken explicitly by the Weyl compensator field and we will not consider it further. 

In the infrared (IR), below the strong-coupling scale $\Lambda$, the theory is described in terms of mesons $M_i^a =  Q^a_\alpha \Qb^\alpha_i$ and $H_{ij} = A_{\alpha\beta}\Qb_i^\alpha\Qb_j^\beta$ as well as baryons $B_n$ and $\bar{B}$. This last baryon exists only for theories with $\Nf \ge 4$ and therefore will not be discussed further in this text. Note that the index on $B_n$ is not a flavor index but instead simply labels the various baryons and must satisfy the constraints i) $\Nc$ and $n$ must be both even or both odd and ii) $n\le \text{min}(\Nc,\Nf)$. For the theories we consider, $n\in \{0,1,2\}$ and schematically $B_n \sim Q^n A^{\frac{\Nc-n}{2}}$ with appropriate contractions with Levi-Civita tensors, as described below. We will denote the portion of the UV global symmetry group in~\cref{eq:UVglobal} that remains in the vacuum as $\mathcal{G}_{vac}(\Nc,\Nf)$.

The proposed analysis was applied to the case of $\Nf=0$ in~\cite{Csaki:2021xhi} by one of the current authors. They found that for $\Nc=2k+1$, the non-supersymmetric theory has massless fermions while there were no massless fermions for $\Nc=2k$. In the present work, we extend the analysis to the cases $\Nf=1,2$ and discover a similar pattern. 

Our analysis will consist of three parts. First, we consider the SUSY theories with an AMSB deformation and work with the UV fields $\{A,Q,\widetilde{Q}\}$ in a weakly-coupled calculation to find stable vacua. This approach will be justified a posteriori by the observation that the vevs are much larger than the strong-coupling scale $\Lambda$. Second, we will argue for the veracity of these vacua via consistency conditions, such as 't Hooft anomaly matching, massless (pseudo-)scalar content, and the supertrace condition
\begin{equation}
    \text{Str}(M^2) = -2 \text{tr}(M_f^2) + \text{tr}(M_b^2) = 0,
\label{eq:supertrace}
\end{equation}
which must be true for any theory with a canonical K\"ahler potential. We will also perform a numerical analysis to confirm our vacua. Thirdly, we will consider the $m\rightarrow \infty$ limit and carry the above statements to non-supersymmetric chiral gauge theories.

The parts of the above analysis are divided amongst the sections of the paper. In~\cref{sec:oneflav}, we specialize to gauge theories with $\Nf=1$ and describe the D-flat directions of the theory as well as the vacuum ansatz for AMSB-deformed gauge theory. In~\cref{sec:twoflav}, we perform an identical analysis for the more complicated case of $\Nf=2$. These results are then mapped to the non-SUSY gauge theories in~\cref{sec:nonsusy}. A numerical analysis to determine the minima and support the ans\"{a}tze of the previous sections is described in~\cref{sec:num}. Finally, in~\cref{sec:conc} we conclude. Various technical details are left for~\cref{sec:charges,sec:masses}.

\begin{table*} 
\def\arraystretch{2}
\begin{tabular}{| c | c | c | c | c | c | c | } 
 \hline
   & $\SU(\Nc)$ & $\SU(\Nf)$ & $\SU(\Nc+\Nf-4)$ & $\U(1)_1$ & $\U(1)_2$ & $\U(1)_R$ \\ [0.5ex] 
 \hline\hline
 $A$ &{\tiny $ \yng(1,1) $}  & 1 & 1 & 0 & -2$\Nf$ & $-\frac{12}{\Nc}$ \\ 
 \hline
 $Q$ & {$\tiny
\Yvcentermath1
\yng(1)$} & {$\tiny
\Yvcentermath1
\yng(1)$} & 1 & 1 & $\Nc-\Nf$ & $2 - \frac{6}{\Nc}$ \\
 \hline
 $\Qb$ & {$\tiny
\Yvcentermath1
\bar{\yng(1)}$} & 1 & {$\tiny
\Yvcentermath1
\yng(1)$} & $\frac{-\Nf}{\Nc+\Nf-4}$ & $\Nf$ & $\frac{6}{\Nc}$  \\
 \hline\hline
 $H$ & 1 & 1 & {\tiny $
\yng(1,1)$} &  $\frac{-2\Nf}{\Nc+\Nf-4}$ & 0 & 0 \\
 \hline
  $M$ & 1 & {$\tiny
\Yvcentermath1
\yng(1)$} & {$\tiny
\Yvcentermath1
\yng(1)$} & $\frac{\Nc-4}{\Nc+\Nf-4}$ & $\Nc$ & 2 \\
 \hline
  $\Bn$ & 1 & \Large\textasteriskcentered & 1 & $n$ & $(n-\Nf)\Nc$ & -4 \\
   \hline
\end{tabular}
\caption{Global charges of the fields in the SUSY gauge theories. The asterisk for $\Bn$ indicates that its representation under $\SU(\Nf)$ changes with $n$: $B_1$ is in the fundamental representation, whereas $B_0$ and $B_2$ are singlets.}
\label{tab:charges}
\end{table*}
\vspace{-3mm}
\section{Gauge Theories with $\Nf=1$}
\label{sec:oneflav}
In this section, we determine exact features of supersymmetric chiral $\SU(\Nc)$ gauge theories with $(\Nc-3)$ fermions in the anti-fundamental representation, one fermion in the fundamental representation, and a single fermion in the antisymmetric representation, with small supersymmetry breaking. We first list the general features of the corresponding SUSY gauge theories. In addition to the gauge-invariant polynomials $H$ and $M$ described above, theories with $N_C$ even (odd) have a baryon $B_0$ ($B_1^a$) given by
\begin{equation}
    \begin{aligned}
        B_0 &= \epsilon^{\alpha_1\alpha_2\cdots \alpha_{\Nc}}A_{\alpha_1\alpha_2}\cdots A_{\alpha_{\Nc-1}\alpha_{\Nc}} ,\qquad \\
        B_1^a &= Q^a_{\alpha_{\Nc}}\epsilon^{\alpha_1\alpha_2\cdots \alpha_{\Nc}}A_{\alpha_1\alpha_2}\cdots A_{\alpha_{\Nc-2}\alpha_{\Nc-1}} .
    \end{aligned}
\label{eq:Nf1baryons}
\end{equation}
Note that even though we write flavor superscripts on $B_1$ and $Q$, they are only relevant for $\Nf >1$. Thus the total number of gauge-invariant polynomial fields is $\binom{\Nc-3}{2} + (\Nc-3) + 1 = \frac{1}{2}(\Nc^2-5\Nc+8)$ for all values of $\Nc$. This matches the number of D-flat directions, which can be seen as follows. At an arbitrary point in the D-flat moduli space, an $\Sp(2)$ subgroup\footnote{We employ the convention such that $\Sp(2)\cong \SU(2)$.} of the gauge group remains unbroken. Thus $(\Nc^2 -1) - (2^2 -1) = \Nc^2-4$ fields are eaten by the broken gauge group. The total number of D-flat directions is then
\begin{equation}
    \binom{\Nc}{2} + \Nc +\Nc(\Nc-3) - (\Nc^2-4) = \frac12 (\Nc^2-5\Nc+8).
\end{equation}
However, due to the large global symmetry group, the number of distinct parameters needed to describe the flat directions is smaller than this value. Let us examine the case of $\Nc =2k+1$ first. Using the $\SU(\Nc-3)$ subgroup of $\mathcal{G}_{UV}(\Nc,1)$, $H_{ij}$ can be taken to have a skew-diagonal form and has $k-1$ distinct parameters. This leaves an $\Sp(2)^{k-1}$ subgroup of $\mathcal{G}_{UV}(\Nc,1)$ unbroken, which in turn can be used to rotate $M$ such that it has only $k-1$ non-zero components. Including the baryon $B_1$, the total number of parameters required is then
\begin{equation}
    N_{\text{flat}}^{(\Nf=1)} = 2\left(k-1 \right) +1 = \Nc -2.
\label{eq:Nf1parameters}
\end{equation}
The same calculation can be done for $\Nc=2k$ with minor differences. In this case, $H$ can be made skew-diagonal except for an additional $0$ entry on the diagonal since $\Nc-3 = 2k-3$ is necessarily odd (see~\cref{eq:NF1evenansatz2}). Thus $H$ is parametrized by $k-2$ distinct values, leaving a  $\SU(2)^{k-2}$ unbroken symmetry group that allows for $k-2$ non-zero values of $M$. Including the baryon $B_0$, we again find 
the same number of parameters as displayed in~\cref{eq:Nf1parameters}

The unbroken $\Sp(2)$ subgroup of the $\SU(\Nc)$ gauge group generates a non-perturbative superpotential of the form
\vspace{-5mm}
\begin{equation}
    W =  \begin{cases} 
\Lambda^{2k+1}/(B_0MH^{(k-2)})^{1/2} & \Nc = 2k \\
      \Lambda^{2k+2}/(B_1\text{Pf}(H))^{1/2} & \Nc=2k+1 \\
   \end{cases}
\label{eq:Nf1superpot}
\end{equation}
where Pf($H$) denotes the Pfaffian of $H_{ij}$. 

In the following subsections we deform these theories utilizing AMSB as encapsulated in~\cref{eq:AMSBpot} and determine the resulting 
vacuum. In doing so, it proves useful to define the following set of matrices:
\begin{equation}
    \mathcal{R}_n := \begin{pmatrix} 0 & \rho^n \\ -\rho^n & 0\end{pmatrix}
\end{equation}
with $\rho\in\mathbb{R}$.
\vspace{-1cm}

\subsection*{Odd Case: $\Nc = 2k+1$}
As discussed in the introduction, our analysis of the AMSB-induced vacuum will originate with the UV fields $\{Q,\Qb, A\}$ as opposed to the IR fields $\{H,M,B\}$. We will see that this approach is justified once the vevs of the squarks are determined and shown to be parametrically larger than $\Lambda$. We begin by describing the D-flat directions. Utilizing the $\SU(\Nc)$ gauge symmetry, we skew-diagonalize $A_{\alpha\beta}$ such that $A = i \sigma_2\, \otimes\, \diag(\alpha_1,\dots,\alpha_{k})\oplus0$. This leaves an $\SU(2)^{k}$ gauge subgroup unbroken which can be used to simplify $Q$ to $Q_\alpha = (\gamma_1, 0, \gamma_2,0,\dots,\gamma_{k-1},0,0,0,\beta)$. Finally, we write $\Qb$ such that its first column resembles $Q$ and the rest fills out the sparse pattern:
\begin{equation}
    \scalebox{0.7}{
    $\Qb = \begin{pmatrix} \delta_1 \\
                            0 & \delta_2 \\
                            \epsilon_1^{(1)} & 0 & \delta_3 \\
                            0 & 0 & 0 & \delta_4 \\
                            \epsilon_1^{(2)} & 0 & \epsilon_2^{(2)} & 0 & \delta_5\\
                            \vdots & \vdots & \vdots & \vdots & \ddots & \ddots \\
                            0 & 0 & 0 & 0 & 0 & \ddots & \delta_{\Nc-5} & \\
                            \epsilon_{1}^{(k-2)} & 0 & \epsilon_{2}^{(k-2)} & 0 & \epsilon_{3}^{(k-2)} & \dots & 0 & \delta_{\Nc-4} \\
                            0 & 0 & 0 & 0 & 0 & \dots & 0 & 0 & \delta_{\Nc-3} \\
                            0 & 0 & 0 & 0 & 0 & \dots & 0 & 0 & 0 \\
                            0 & 0 & 0 & 0 & 0 & \dots & 0 & 0 & 0 \\
                            \eta_1 & 0 & \eta_2 & 0 & \eta_3 & \dots & 0 & \eta_{k-1} & 0 
              \end{pmatrix}$
}\fstop
\end{equation}
Substituting the above into the D-flat condition~\cref{eq:DFlat}, we obtain the following constraint equations:
\vspace{-2mm}
\begin{align}
    \sigma_{k}^2 &= \sigma_a^2 + \gamma_a^2 - \delta_{2a - 1}^2 - \sum_{b=1}^{a-1} \epsilon_b^{(a)} = \sigma_a^2 - \delta_{2a}^2, \\
    &\hspace{47mm}\text{ for all } a < k \nonumber\\
    \sigma_k^2  &= \beta^2 - \sum_{b=1}^{k-1} \eta_b^2 \\
    0 &= \gamma_a \gamma_b - \delta_{2a-1} \epsilon_{a}^{(b-1)} - \sum_{c = 1}^{a - 1} \epsilon_c^{(a - 1)} \epsilon_c^{(b - 1)}, \\
    &\hspace{47mm}\text{ for all } a < k \,\, \& \,\, b > a, \nonumber\\
    0 &= \gamma_a \beta - \eta_a \delta_{2a - 1} - \sum_{b=1}^{a - 1} \eta_b \epsilon_b^{a - 1} \hspace{0.5mm}\text{  for all } a \leq k.
\end{align}
After satisfying these constraints, the number of free parameters is $\Nc -2$, as expected. The resulting parametrizations for the fields can then be substituted into the scalar potential~\cref{eq:AMSBpot} and vacua can be determined via a multi-dimensional minimization procedure. We performed this task explicitly for low values of $\Nc$, but the problem quickly becomes intractable due to the huge number of parameters. However, the low-$\Nc$ cases revealed a universal ansatz, which we now use for all $\Nc$. Our vacuum ansatz takes the form
\vspace{-2mm}
\begin{equation}
\begin{aligned}
     \langle A_{\alpha\beta}\rangle &= \frac{1}{\sqrt{2}}\begin{bmatrix} 2\mathcal{R}_1 & 0 & \dots & \dots & 0\\
    0 &  \ddots & \ddots & & \vdots\\
    \vdots & \ddots & 2\mathcal{R}_1 & \ddots & \vdots\\
    \vdots & & \ddots & \sqrt{2}\mathcal{R}_1 & 0 \\
    0 & \dots & \dots & 0 &  0\\
    \end{bmatrix},\\
    \langle Q_\alpha\rangle &= \sqrt{2}\begin{bmatrix}0,0,0,\cdots,0, \rho\end{bmatrix}^T,\\ 
    \langle\widetilde{Q}^\alpha_i\rangle &= \sqrt{2}\begin{bmatrix}\rho & & 0 & 0 & 0 & 0\\
     & \ddots &  & 0 & 0 & 0\\
    0 &  &\rho & 0 & 0 & 0\end{bmatrix}\fstop
\end{aligned}
\label{eq:NF1oddansatz}
\end{equation}
where $\rho$ is an as-yet undetermined real number. We see here explicitly the existence of an unbroken $\Sp(2)$ gauge subgroup that supports the gaugino condensate superpotential. From the ansatz~\cref{eq:NF1oddansatz}, the gauge-invariant polynomials take the form:
\begin{equation}
\begin{aligned}
    \langle B_1\rangle &= k!\frac{(2\sqrt{2}\rho)^{k+1}}{2\sqrt{2}}, \\
    \langle H\rangle &= 2\sqrt{2}\begin{bmatrix}
                \mathcal{R}_3 & 0 & \cdots & 0\\
                0 & \mathcal{R}_3 & \dots & 0\\
                \vdots & & & \vdots\\
                0 & \cdots & \cdots & \mathcal{R}_3
            \end{bmatrix},\\
            \langle M\rangle &= 0,
\end{aligned}
\label{eq:NF1oddansatz2}
\end{equation}
such that
\begin{equation}
    \langle \text{Pf}(H)\rangle  = (2\sqrt{2}\rho^3)^{k-1} \fstop
\end{equation}
Substituting the ansatz into the lower superpotential of~\cref{eq:Nf1superpot}, we find
\begin{equation}
    \langle W\rangle  = (k!)^{-1/2}2^{-(6k-3)/4}\Lambda^{2k+2}\rho^{-(2k-1)}.
\end{equation}
The total potential from~\cref{eq:AMSBpot} , including AMSB contributions, is then
\begin{equation}
    \langle V \rangle = \frac{(\Nc-2)W^2}{4\rho^2} - m(\Nc+1)W.
\end{equation}
We then determine a minimum exists with 
\begin{equation}
    \rho = \left( \frac{(\Nc-1) \Lambda^{\Nc +1}}{(2^{(3\Nc+2)/2}k!)^{1/2}(\Nc +1)m}\right)^{1/\Nc}.
\label{eq:Nf1oddvev}
\end{equation}
We can now justify our approach in performing the above analysis with the UV fields. For $m\ll \Lambda$, the value of $\rho$ is parametrically larger than $\Lambda$. Thus any calculation performed purely with the IR fields would necessarily be suspect, and a thorough analysis requires the use of the UV fields, as we have done here.

We now consider the particle spectrum in this vacuum. From the form of the superpotential in~\cref{eq:Nf1superpot}, we see that the fermionic components of the mesons $M$ will remain massless regardless of the vevs of the other fields. The masses of all other fermions, scalars and pseudoscalars are shown in \cref{tab:masses} in~\cref{sec:masses}. The massless fermions allow for the existence of non-trivial 't Hooft anomalies of global symmetries, which must match between the UV and IR for consistency. The form of~\cref{eq:NF1oddansatz2} indicates that the vacuum has an unbroken global symmetry group 
\begin{equation}
    \mathcal{G}_{vac}(2k+1,1) = \Sp(N_C-3)\times \U(1)_2.
\label{eq:Nf1oddglobal}
\end{equation}
Here $\U(1)_2$ is the same Abelian global symmetry present in~\cref{eq:UVglobal}. From the charge assignments of the $A$, $Q$, and $\widetilde{Q}$ fields in~\cref{tab:charges} in~\cref{sec:charges}, we have the following non-zero 't Hooft anomaly coefficients:
\begin{equation}
    \begin{aligned}
        \text{Tr}[\U(1)_2] &= \Nc(\Nc-3),\\
        \text{Tr}[(\U(1)_2)^3] &= \Nc^3(\Nc-3),\\
        \text{Tr}[\U(1)_2 \, \Sp(\Nc-3)^2] &= \frac12 \Nc.
    \end{aligned}
\label{eq:OddF1Hooft}
\end{equation}
Referring to ~\cref{tab:charges} once again, we see that these coefficients are also obtained by the fermionic components of the $M$, thereby verifying the local anomaly matching. Since $\pi_4(\Sp(n))\cong \mathbb{Z}_2$ for all $n\in\mathbb{Z}$, we must also check the matching of a potential Witten anomaly~\cite{Witten:1982fp} for the global $\Sp(\Nc-3)$ symmetry. This anomaly is present in the IR as there is a single massless fermion charged under $\Sp(\Nc-3)$. However, this is matched by an analogous global anomaly in the UV that arises because there are $\Nc =1$ mod 2 fermions charged under $\Sp(\Nc-3)$. Furthermore, in the cases we checked explicitly, we found the the number of massless scalars matches precisely with the symmetry breaking pattern discussed above, and the supertrace condition~\cref{eq:supertrace} is satisfied. 

In~\cref{sec:nonsusy} we will carry the above results over to non-SUSY gauge theories. For the moment, we move onto the case of even $\Nc$ theories.
\vspace{-2mm}
\subsection*{Even Case: $\Nc=2k$}
As in the odd $\Nc$ case above, we begin by skew-diagonalizing $A_{\alpha\beta}$ with the $\SU(\Nc)$ gauge symmetry such that $A = i \sigma_2 \otimes \diag(\alpha_1,\dots,\alpha_{k-2},\beta,\beta)$. This leaves an $\SU(2)^{k-2} \times \Sp(4)$ gauge subgroup unbroken which can be used to simplify  the fundamental fields as $Q_\alpha = (\gamma_1, 0, \gamma_2,0,\dots,\gamma_{k-1},0,0,0)$. Now, we set $\Qb_{2k-3}^\alpha = Q_\alpha$ so their contributions to~\cref{eq:DFlat} cancel. The remaining elements of $\Qb$ are given by
$\Qb = \diag(\delta_1,\delta_1,\dots,\delta_{k-2},\delta_{k-2})$ for the upper $(2k-4) \times (2k-4)$ part of the matrix, while the lower $4 \times (2k-4)$ vanishes. 
Thus, the D-flatness condition~\cref{eq:DFlat} reduces to $|\alpha_1|^2 - |\delta_1|^2 =...=|\alpha_{k-2}|^2 - |\delta_{k-2}|^2 = |\beta|^2$. Thus the independent parameters are the $k-1$ $\{\gamma_i\}$, $k-2$ $\{\delta_i\}$, and the $\beta$, for a total of $2k-2 = \Nc-2$, as predicted in the previous section. Also as above, we utilize the following ansatz to search for a vacuum:
\begin{equation}
\begin{aligned}
    \langle\widetilde{Q}^\alpha_i\rangle &= \sqrt{2}\begin{bmatrix}\rho & & 0 & 0 & 0 & 0\\
     & \ddots &  & 0 & 0 & 0\\
    0 &  &\rho & 0 & 0 & 0\end{bmatrix},\\
     \langle A_{\alpha\beta}\rangle &= \frac{1}{\sqrt{2}}\begin{bmatrix} 2\mathcal{R}_1 & 0 & \dots & \dots & 0\\
    0 &  \ddots & \ddots &  & \vdots\\
    \vdots & \ddots & 2\mathcal{R}_1 & \ddots & \vdots\\
    \vdots &  & \ddots & \sqrt{2}\mathcal{R}_1 & 0 \\
    0 & \dots & \dots & 0 &  \sqrt{2}\mathcal{R}_1 \\
    \end{bmatrix},\\
    \langle Q_\alpha\rangle  &= \sqrt{2}\begin{bmatrix}
        0,0,\cdots,0,\rho,0,0,0
    \end{bmatrix}^T.
\end{aligned}
\label{eq:NF1evenansatz}
\end{equation}
Again we explicitly see the unbroken $\Sp(2)$ subgroup of the gauge group from this parametrization, as required for the existence of the non-perturbative superpotential in~\cref{eq:Nf1superpot}. The ansatz in~\cref{eq:NF1evenansatz} gives compact expressions for the gauge-invariant polynomials:
\begin{equation}
\begin{aligned}
    \langle B_0\rangle  &= \frac{1}{2}k!(2\sqrt{2}\rho)^{k}, \\
    \langle H_{ij}\rangle &= 2 \sqrt{2}\begin{bmatrix}\mathcal{R}_3 &0 & \dots & 0, \\
    0 &\ddots & \ddots & \vdots \\ 
    \vdots & \ddots & \mathcal{R}_3 & 0 \\ 
    0 & \dots & 0 & 0   \end{bmatrix}, \\
    \langle M_i\rangle  &=  \begin{bmatrix} 0,0, \cdots ,0, 2 \rho^2 \end{bmatrix} \fstop
\end{aligned}
\label{eq:NF1evenansatz2}
\end{equation}
While the superpotential takes a similarly simple form:
\begin{align}
    \langle W \rangle = \left(k! (k-2)!\right)^{-1/2}\sqrt{2}(2\rho)^{2-2k}\Lambda^{2k+1}.
\end{align}
The total potential from~\cref{eq:AMSBpot} , including AMSB contributions, is then
\begin{equation}
    \langle V \rangle = \frac{(\Nc -2)W^2}{4\rho^2} - m(\Nc +1)W.
\end{equation}
At the minima, we find 
\begin{equation}
    \rho = \left( \frac{(\Nc -1) \Lambda^{\Nc +1}}{(2^{2\Nc-3}k!(k-2)!)^{1/2}(\Nc +1)m}\right)^{1/\Nc}.
\label{eq:Nf1evenvev}
\end{equation}

The global symmetry group~\cref{eq:UVglobal} breaks to
\begin{equation}
    \mathcal{G}_{vac}(2k,1) = \Sp(\Nc-4)\times \U(1)_3
\label{eq:Nf1evenglobal}
\end{equation}
in the vacuum described by~\cref{eq:NF1evenansatz,eq:Nf1evenvev}. Here $\U(1)_3$ is a linear combination of two $\U(1)$'s in $\mathcal{G}_{UV}$: 
\begin{equation}
    Q_3 = Q_{\SU(\Nc-3)} + (\Nc-3)Q_1.
\label{eq:Nf1NcEvenQ}
\end{equation}
Where $Q_1$ is the $\U(1)_1$ charge displayed in~\cref{tab:charges} and $Q_{\SU(\Nc-3)}$ is determined by the $\SU(\Nc-3)$ Cartan subalgebra generator $\diag (1,1,1,\cdots,-(\Nc-4))$. The $\U(1)_3$ charges of the UV and IR fields are displayed in \cref{tab:charges2}.

In the vacuum described above, we find no massless fermions. We can see this directly with the form of the superpotential ~\cref{eq:Nf1superpot} as follows. The combination of $M$ and $H$ is contracted with the fully antisymmetric tensor of $\text{SU}(N_C+N_F-4)=\text{SU}(N_C-3)$. Looking at the expression of gauge invariants in the vacuum state in ~\cref{eq:NF1evenansatz2}, we can think of the globally invariant expression of $M$ and $H$ as the last component of $M$ times the Pfaffian of $H$ without the last row and column to make the restriction of $H$ an even dimensional matrix so that the Pfaffian is well-defined. To yield a non-vanishing term the fermion mass matrix, we can either take derivatives with respect to the non-zero components in ~\cref{eq:NF1evenansatz2}, or differentiate two pairs of components of $H$ or a pair of components of $H$ and one component of $M$ that leave neighboring indices $i,i+1$ with $i$ an odd number leftover to contract with the antisymmetric tensor. As a result, the fermion mass matrix block diagonalizes into $\textbf{k}\oplus\textbf{2}\oplus\cdots\oplus\textbf{2}$, where all the $2\times2$ blocks are anti-diagonal matrices that produce non-zero mass eigenstates, and the $k\times k$ block (the non-zero components of the gauge invariants) have no massless eigenvalues, since the determinant of this sub-block of the fermion mass matrix is
\begin{equation}
    \det \langle W^{\prime\prime}|_{k\times k}\rangle=\frac{N_C+4}{2^{(N_C+4)/2}B_0^2H^{N_C-4}M^2},
\end{equation}
where $B_0,H,M$ here refer to the non-zero entries of the gauge invariants for the vacuum ~\cref{eq:NF1evenansatz2}. The complete list of masses of the fermions, scalars and pseudoscalars are shown in \cref{tab:masses}.

Congruent with the non-existence of massless fermions, all of the global anomalies of~\cref{eq:Nf1evenglobal} vanish in the UV:
\vspace{-2mm}
\begin{equation}
\begin{aligned}
    &\Tr[\U(1)_3]: 0 + \Nc(\Nc-3)+\Nc(3-\Nc)=0,\\
    &\Tr[(\U(1)_3)^3]: 0 + \Nc(\Nc-3)^3+\Nc(3-\Nc)^3=0 ,\\
    &\Tr[\U(1)_3\,\Sp(\Nc-4)^2]: 0 \fstop
\end{aligned}
\end{equation}
The last of these trivially vanishes because the fields charged under the $\Sp(\Nc - 4)$ subgroup include only $\widetilde{Q}$ and elements of $H_{ij}$ for $i \neq \Nc - 3$, while the $\U(1)_3$ is only nonzero for $i = \Nc - 3$ for those fields. As in the odd $\Nc$ case above, we must also consider global anomalies of $\Sp(\Nc - 4)$. In the IR we have no massless fermions and thus no global anomaly. The UV also lacks a global anomaly as there are $\Nc=0$ mod 2 charged fermions.
Furthermore, in the set of theories we explicitly checked, the numbers of massless scalars matches the numbers of broken generators, and we find that the supertrace sum rule in~\cref{eq:supertrace} holds. The preceding statements indicate that the proposed vacuum is indeed valid. 
\vspace{-3mm}
\section{Gauge Theories with $\Nf=2$}
\label{sec:twoflav}
We now move onto the case $\Nf=2$, which features an enlarged flavor symmetry group compared to the previous theories. As before, we focus on the supersymmetric theories first. The gauge-invariant polynomials include the $H$ and $M$ fields as before, but there are two important differences compared to the previous $\Nf=1$ case. First, for $\Nc=2k+1$, we again have the baryon field $B_1^a$ from~\cref{eq:Nf1baryons}, but it now transforms in the fundamental representation of the global $\SU(\Nf)=\SU(2)$ factor of~\cref{eq:UVglobal}. Secondly, for $\Nc=2k$, we supplement $B_0$ of~\cref{eq:Nf1baryons} with an additional invariant:
\vspace{-2mm}
\begin{equation}
\label{eq:Nf2Binv}
    B_2 = \varepsilon_{ab}Q^a_{\alpha_1} Q^b_{\alpha_2}\varepsilon^{\alpha_1\alpha_2\cdots\alpha_{2k}}A_{\alpha_3\alpha_4}\cdots A_{\alpha_{2k-1}\alpha_{2k}}\fstop
\end{equation}

The total number of gauge-invariant polynomials is now $\binom{\Nc-2}{2}+2(\Nc-2)+2 = \frac12 (\Nc^2-\Nc+2)$ for all values of $\Nc$. This matches the D-flat directions, where now we expect that the full gauge group is broken at an arbitrary point in the moduli space. Then the number of moduli is 
\vspace{-2mm}
\begin{equation}
    \binom{\Nc}{2}+2\Nc+ \Nc(\Nc-2)- (\Nc^2-1)=\frac12 (\Nc^2-\Nc+2)\fstop
\end{equation}

Here we will start with the case $\Nc=2k+1$. The global $\SU(\Nc-2)$ symmetry of $\mathcal{G}_{UV}(\Nc,2)$ allows us to rotate $H$ into a skew-diagonal form with an extra 0 entry on the diagonal, while retaining an $\Sp(2)^{(\Nc-3)/2}$ subgroup unbroken. Then $H$ is determined by $(\Nc-2)/2$ distinct eigenvalues, and the leftover $\Sp(2)^{(\Nc-2)/2}$ allows us to parametrize $M$ with only $\frac{\Nc-3}{2}+1+\Nc-2 = \frac12 (3\Nc-5)$ nonzero values. We use the global $\SU(2)$ of~\cref{eq:UVglobal} to enforce a single non-vanishing entry of $B_1^a$, leaving the required number of parameters to be
\vspace{-2mm}
\begin{equation}
    N_{flat}^{(\Nf=2)} = \frac{\Nc - 3}{2} + \frac12 (3\Nc-5) + 1  = 2\Nc-3.
\label{eq:Nf2parameters}
\end{equation}

For $\Nc=2k+1$, we can again skew-diagonalize $H$, but now with no extra 0 entry. This form of $H$ gives $(\Nc-2)/2$ distinct eigenvalues and an unbroken $\Sp(2)^{(\Nc-3)/2}$ subgroup of $\SU(\Nc-2)$. We can then use this subgroup as well as the global $\SU(2)$ to leave $\frac{\Nc-2}{2}+(\Nc-2)-1 =\frac{3\Nc}{2}-4$ non-zero entries of $M$. Including the two baryons $B_0$ and $B_2$, we find the same number of parameters as found in~\cref{eq:Nf2parameters}:
\begin{equation}
    N_{flat}^{(\Nf=2)} = \frac{\Nc-2}{2}+\frac{3\Nc}{2}-4+2 = 2\Nc-3 .
\label{eq:Nf2parameters2}
\end{equation}
The gauge group is fully broken, but a non-perturbative superpotential is still generated by instantons with the form~\cite{Poppitz:1995fh,Pouliot:1995me}.
\begin{align}
    W &= \nonumber \\ & \begin{cases} 
      \Lambda^{4k+3}(\Bi MH^{k-1}) & \Nc=2k+1 \\
      \Lambda^{4k+1}/[\Bii\text{Pf}(H) - \frac{k - 1}{k}\Bo M^2 H^{k-2}] & \Nc=2k \\
   \end{cases}
\label{eq:Nf2super}
\end{align}

We now deform these SUSY theories via AMSB and search for vacua. 
\vspace{-5mm}
\subsection*{Odd Case: $\Nc=2k+1$}
As in the $\Nf=1$ theories above, we will search for vacua using the UV fields as opposed to the IR fields. We begin by describing the D-flat parametrizations. Rotations from the $\SU(\Nc)$ gauge group allow us to cast $A$ in the now-familiar skew diagonal form $A= i\sigma_2 \otimes \diag(\sigma_1,...,\sigma_k) \oplus 0$. The leftover $\Sp(2)^k$ allows us to remove a number of entries in $Q^a_\alpha$ such that it takes the form
\begin{equation}
    Q^a_\alpha = \begin{bmatrix}
        \gamma_1 & \gamma_2 & \gamma_3 &\gamma_4 & \cdots &\gamma_{2k-1} &\gamma_{2k}& \gamma_{2k+1}\\
        \beta_1 & 0 & \beta_2 & 0 &\cdots & \beta_k & 0 & \sigma_1
    \end{bmatrix}^T.
\end{equation}
Note that we have made one choice beyond the $\Sp(2)^k$ rotations to abbreviate the analysis -- the last entry of the second column has been chosen to match one of the eigenvalues of $A$. Finally, we write $\Qb^\alpha_i$ such that its last two columns nearly match those of $Q^a_\alpha$ 
\begin{equation}
    \scalebox{0.8}{
    $\Qb^\alpha_i =\begin{bmatrix}
        0 & 0 & 0 &\cdots & 0 & \beta_1 &\gamma_1  \\
        0 & 0 & 0 &  \cdots & 0 & 0 & \gamma_2 \\
        0 & \epsilon^{(1)}_2 & 0 &  \cdots & 0 & \beta_2 + \delta_1 & \gamma_3 \\
        \epsilon_1^{(1)}& 0 & \epsilon^{(1)}_3 &  \cdots & \epsilon^{(1)}_{2k-1} &0 & \gamma_4  \\
       0 & \epsilon_2^{(2)} & 0 &  \cdots & 0 & \beta_3 +\delta_2 &\gamma_5  \\
        \epsilon_1^{(2)} & 0  & \epsilon_3^{(2)} &  \cdots & \epsilon_{2k-1}^{(2)} & 0 & \gamma_6 \\
        \vdots & \vdots & \vdots &  \vdots & \vdots & \vdots & \vdots\\
        0 & \epsilon_2^{(k-1)} &0 &  \cdots & 0 & \beta_k+\delta_{k-1} &\gamma_{2k-1} \\
         \epsilon_1^{(k-1)} & 0 & \epsilon_3^{(k-1)} &  \cdots & \epsilon_{2k-1}^{(k-1)} & 0 &  \gamma_{2k} \\
        0 & 0 & 0 & \cdots & 0 & 0 & \gamma_{2k+1} 
    \end{bmatrix}$
    }.
\end{equation}

Just as in \cref{sec:oneflav}, we also enforce the final D-flat constraints, which straightforwardly resemble those from before. For these models, our ansatz for the vacuum is 
\vspace{-2mm}
\begin{equation}
\begin{aligned}
     \langle A_{\alpha\beta}\rangle &= \frac{1}{\sqrt{2}}\begin{bmatrix} \sqrt{2}\mathcal{R}_1 & 0 & \dots & \dots & 0\\
    0 &  2\mathcal{R}_1 & \ddots & & \vdots\\
    \vdots & \ddots & \ddots & \ddots & \vdots\\
    \vdots & & \ddots & 2\mathcal{R}_1 & 0 \\
    0 & \dots & \dots & 0 &  0\\
    \end{bmatrix},\\
    \langle Q_\alpha^a\rangle 
    &= \sqrt{2}\begin{bmatrix}0 & \rho & 0 & 0 & \cdots & 0 & 0\\ 0 &  0 & 0 &0 & \cdots &0 & \rho \end{bmatrix}^T,\\
    \langle \widetilde{Q}^\alpha_i\rangle &= \sqrt{2}
    \begin{bmatrix}
    0 & 0 & 0 & 0 & \dots & 0 & \rho & 0 \\ 
    0 & 0 & 0 & 0 & \iddots & -\rho & 0 & 0 \\ 
    \vdots & \vdots & \vdots & \iddots & \iddots & \iddots & \vdots & \vdots \\ 
    0 & 0 & 0 & \rho & \iddots & 0 & 0 & 0 \\ 
    0 & 0 & -\rho & 0 & \dots & 0 & 0 & 0 \\ 
    0 & \rho & 0 & 0 & \dots & 0 & 0 & 0
    \end{bmatrix}.
\end{aligned}
\label{eq:NF2oddansatz}
\end{equation}
These give compact expressions for the gauge invariant polynomials
\vspace{-2mm}
\begin{equation}
\begin{aligned}
    \langle B_1^a \rangle &= k!(2\sqrt{2})^{k} \rho^{k+1}\begin{bmatrix}
        0, 1
    \end{bmatrix},\\
    \langle H_{ij} \rangle &= 2\sqrt{2}\begin{bmatrix}
               0 & 0 & \dots & 0 & 0 \\
                0 &  \mathcal{R}_3 & \ddots & \vdots & 0 \\
                \vdots & \ddots & \ddots & 0 & \vdots \\
                0 & \dots & 0 & \mathcal{R}_3 & 0 \\
                0 & 0 & \dots & 0 &  \mathcal{R}_3 
            \end{bmatrix}, \\
    \langle M^a_i \rangle 
         &= 2 \begin{bmatrix}\rho^2 & 0 &0 &\cdots & 0 & 0\\
         0 & 0 & 0 &\cdots &0 &0\end{bmatrix}^T.
\end{aligned}
\label{eq:NF2oddansatz2}
\end{equation}
The superpotential is then
\vspace{-2mm}
\begin{equation}
    \langle W \rangle = \frac{2\sqrt{2}}{k!(k-1)!}\frac{\Lambda^{4k+3}}{(2\rho)^{4k}},
\end{equation}
which gives the the F-term scalar potential and AMSB contribution as  
\vspace{-2mm}
\begin{equation}
    \langle V \rangle = \frac{(\Nc-1) W^2}{\rho^2} - 2m(2\Nc +1)W.
\end{equation}
At the minima, we find that $\rho$ is 
\vspace{-2mm}
\begin{align}
    \rho = \left( \frac{(2\Nc -1) \Lambda^{2\Nc +1}}{2^{(4\Nc-5)/2}k!(k-1)!(2\Nc +1)m}\right)^{1/2\Nc}.
\end{align}
We note that this result once again reassures us that the UV approach was correct. The complete list of masses of the fermions, scalars and pseudoscalars are shown in \cref{tab:masses}. The vevs in~\cref{eq:NF2oddansatz2} indicate that the unbroken global symmetry group of the vacuum is
\begin{equation}
    \mathcal{G}_{vac}(2k+1,2) = \Sp(N_C-3)\times \U(1)_4 \times \U(1)_5,
\label{eq:Nf2oddglobal}
\end{equation}
where the two new global $\U(1)$ charges are 
\vspace{-2mm}
\begin{equation}
    \begin{aligned}
        Q_4 &= \widetilde{Q}_{\SU(\Nc-2)} + \frac{\Nc-2}{2}Q_1 + \frac
    {\Nc-2}{2\Nc}Q_2,\\
    Q_5 &= -NQ_{\SU(2)}+ Q_2.
    \end{aligned}
\end{equation}
Here $Q_{\SU(2)}$ is the charge under the single Cartan subalgebra generator $\sigma_3 = \diag(1,-1)$ of the global $\SU(2)$ factor in~\cref{eq:UVglobal} and $\widetilde{Q}_{\SU(\Nc-2)}$ is $\diag(-(\Nc-3),1,1,\cdots,1)$. We now verify 't Hooft anomaly matching -- for the UV fields
\begin{equation}
\begin{aligned}
\text{Tr}[\U(1)_4] &= (\Nc-2)(\Nc-3),\\
\text{Tr}[(\U(1)_4)^3] &= (\Nc-3)(\Nc-2)^3,\\
\Tr[\U(1)_4\,\Sp(\Nc-3)^2] &= \frac12 (\Nc-2).
\end{aligned}
\end{equation}
and
\vspace{-2mm}
\begin{equation}
\begin{aligned}
\text{Tr}[\U(1)_5] &= 2\Nc (\Nc-3),\\
\text{Tr}[(\U(1)_5)^3] &= 8\Nc^3 (\Nc-3),\\
\Tr[\U(1)_5\Sp(\Nc-3)^2] &= \frac12 \times2\Nc.
\end{aligned}
\end{equation}
There are also mixed anomalies:
\begin{equation}
\begin{aligned}
\text{Tr}[(\U(1)_4)^2 \, \U(1)_5] &=2\Nc (\Nc-3)(\Nc-2)^2, \\
\text{Tr}[\U(1)_4 \, (\U(1)_5)^2] &= 4\Nc^2(\Nc-2)(\Nc-3) .
\end{aligned}
\end{equation}
Finally, there is also a Witten anomaly for the non-Abelian $\Sp(\Nc-3)$ of~\cref{eq:Nf2oddglobal} as the number of $\Qb$ fermions charged under this global group factor is $\Nc = 1$ mod 2. The above  non-zero anomalies imply that there must be massless fermions in our vacuum that match the above coefficients. The existence of massless fermions can be confirmed by calculations on a case-by-case basis, but a simple argument proves their existence for all $\Nc$. In our vacuum determined by~\cref{eq:NF2oddansatz2}, the theories will have $\Nc-3$ massless fermions, which can be identified with the superpartners of $M^2_{i}$ with $i=2,...,\Nc-2$. A simple application of~\cref{eq:NF2oddansatz2} to the fermion mass matrix reveals this fact. All of the second order derivatives that include these fields vanish because they involve either $H_{i,1}$ or $B_1^1$, which vanish in the vacuum. This is true for all $\Nc$. For the charge assignments in~\cref{tab:charges2}, one finds that these fermions match the anomaly coefficients as well as the Witten anomaly above.
\vspace{-5mm}
\subsection*{Even Case: $\Nc=2k$}
The UV field analysis and D-flat parametrization for this case is quite similar to the odd $\Nc$ case above. Rotations from the $\SU(\Nc)$ gauge group allow us to cast $A$ in the skew diagonal form $A= i\sigma_2 \otimes \diag(\sigma_1,...,\sigma_k)$. The leftover $\Sp(2)^k$ allows us to remove a number of entries in $Q^a_\alpha$ such that it takes the form
\vspace{-2mm}
\begin{equation}
    Q^a_\alpha = \begin{bmatrix}
        \gamma_1 & \gamma_2 & \gamma_3 &\gamma_4 & \cdots &\gamma_{2k-1} &\gamma_{2k}\\
        \beta_1 & 0 & \beta_2 & 0 &\cdots & \beta_k & 0
    \end{bmatrix}^T.
\end{equation}
Finally, we write $\Qb^\alpha_i$ such that its last two columns nearly match those of $Q^a_\alpha$ 
\vspace{-2mm}
\begin{equation}
    \scalebox{0.9}{
    $\Qb^\alpha_i =\begin{bmatrix}
        0 & \cdots & 0 & 0 & 0 &\delta_1 &\gamma_1  \\
        0 & \cdots & 0 & 0 & 0 & 0 & \gamma_2 \\
        0 & \cdots & 0 & 0 & 0 & \delta_2 & \gamma_3 \\
        0 & \cdots & \epsilon^{(1)}_3 & 0 & \epsilon_1^{(1)} & 0 & \gamma_4  \\
       \epsilon_{2k-4}^{(2)} & \cdots & 0 & \epsilon_2^{(2)} & 0 & \delta_3 &\gamma_5  \\
        0 & \cdots & \epsilon_3^{(2)} & 0 & \epsilon_1^{(2)} & 0 & \gamma_6 \\
        \vdots & \vdots & \vdots & \vdots & \vdots &\vdots & \vdots\\
        \epsilon_{2k-4}^{(k-2)} &  \cdots & 0 & \epsilon_2^{(k-2)} & 0 & \delta_{k-1} &\gamma_{2k-3} \\
        0 &  \cdots & \epsilon_3^{(k-2)} & 0 & \epsilon_1^{(k-2)} & 0 & \gamma_{2k-2}\\
        \epsilon_{2k-4}^{(k-1)} &  \cdots & 0 & \epsilon_2^{(k-1)} & 0 & 0 &\gamma_{2k-1} \\
        0 &  \cdots & 0 & 0 & 0 & 0 & \gamma_{2k} 
    \end{bmatrix}$
    }.
\end{equation}
From here we can enforce the D-flatness conditions in~\cref{eq:DFlat}. This analysis for the low-$\Nc$ cases gives us insight to construct the following ansatz:
\vspace{-2mm}
\begin{equation}
\begin{aligned}
     \langle A_{\alpha\beta} \rangle &= \frac{1}{\sqrt{2}}\begin{bmatrix} \sqrt{2}\mathcal{R}_1 & 0 & \dots & \dots & 0\\
    0 &  2\mathcal{R}_1 & \ddots & & \vdots\\
    \vdots & \ddots & \ddots & \ddots & \vdots\\
    \vdots & & \ddots & 2\mathcal{R}_1 & 0 \\
    0 & \dots & \dots & 0 &  \sqrt{2}\mathcal{R}_1
    \end{bmatrix},\\
    \langle Q_\alpha^a \rangle &= \sqrt{2}\begin{bmatrix}0 & \rho & 0 & \dots & 0 & 0 \\ 
    0 & 0 & \dots & 0 & \rho & 0 \end{bmatrix}^T,\\
    \langle \widetilde{Q}^\alpha_i \rangle &= \sqrt{2}
    \begin{bmatrix}
    0 & 0 & 0 & 0 & \dots & 0 & \rho & 0 & 0 \\ 
    0 & 0 & 0 & 0 & \iddots & -\rho & 0 & 0 & 0 \\ 
    \vdots & \vdots & \vdots & \iddots & \iddots & \iddots & \vdots & \vdots & \vdots \\ 
    0 & 0 & 0 & \rho & \iddots & 0 & 0 & 0 & 0 \\ 
    0 & 0 & -\rho & 0 & \dots & 0 & 0 & 0 & 0 \\ 
    0 & \rho & 0 & 0 & \dots & 0 & 0 & 0 & 0\\
    0 & 0 & 0 & 0 & \dots & 0 & 0 & \rho & 0 
    \end{bmatrix}.
\end{aligned}
\label{eq:NF2evenansatz_uv}
\end{equation}
These result in the following expressions for the gauge invariant polynomials
\vspace{-2mm}
\begin{equation}
\begin{aligned}
    \langle B_0 \rangle &= \frac{1}{2}k!(2\sqrt{2}\rho)^{k}, \\
    \langle B_2 \rangle &= 0 ,\\
    \langle H_{ij} \rangle &= 2\sqrt{2}\begin{bmatrix}
                \mathcal{R}_3 & 0 & \dots & 0 & 0 & 0\\
                0 & \ddots & \ddots & \vdots & 0 & 0 \\
                \vdots & \ddots & \ddots & 0 & \vdots & \vdots \\
                0 & \dots & 0 & \mathcal{R}_3 & 0 & 0\\
                0 & 0 & \dots & 0 & 0 & 0\\
                0 & 0 & \dots & 0 & 0 & 0
            \end{bmatrix}, \\
    \langle M^a_i \rangle &= 2 \begin{bmatrix}
                0 & \dots & 0 & \rho^2 & 0 \\
                0 & \dots & 0 & 0 & \rho^2
                \end{bmatrix}^T.
\end{aligned}
\label{eq:NF2evenansatz2}
\end{equation}
In terms of the vev, the superpotential takes the form
\begin{equation}
    \langle W \rangle = \frac{2\Lambda^{4k + 1}}{((k-1)!)^2(2\rho)^{2(2k - 1)}},
\end{equation}
which gives the the F-term scalar potential and AMSB contribution as  
\begin{equation}
    \langle V \rangle = \frac{(\Nc -1)W^2}{\rho^2} -2m(2\Nc +1)W,
\end{equation}
where the vev can now be solved for as:
\begin{equation}
    \rho = \left(\frac{(2\Nc - 1)\Lambda^{2\Nc+1}}{2^{2(\Nc - 1)}((k-1)!)^2(2\Nc + 1)m}\right)^{1/2\Nc}.
\end{equation}

The unbroken subgroup of the global symmetry group in this vacuum is
\begin{equation}
    \mathcal{G}_{vac}(2k,2) = \SU(2)_D \times \Sp(\Nc - 4) \times \U(1)_6,
\end{equation}
where $\SU(2)_D$ is identified as the diagonal subgroup of $\SU(2) \times \SU(2)$, originating from the $\SU(\Nf)$ and leftover symmetry from the $\SU(\Nc - 2)$.
The new global $\U(1)_6$ charge is 
\begin{equation}
    Q_6 = \widetilde{Q}_{1,\SU(\Nc-2)} + \left( \frac{\Nc - 4}{\Nc - 2}\right)\widetilde{Q}_{2,\SU(\Nc-2)} + (\Nc - 3) Q_1.
\end{equation}
Here, $\widetilde{Q}_{1,\SU(\Nc-2)}$ and $\widetilde{Q}_{1,\SU(\Nc-2)}$ are defined respectively as two of the Cartan generators of the $\SU(\Nc-2)$ subgroup, $\text{diag}(1, \dots, 1, 3 - \Nc)$ and $\text{diag}(1, \dots, 1, 4 - \Nc, 0)$. The vacuum given above produces no massless fermions, which can be seen by block diagonalizing the fermion mass matrix similarly to the $N_F=1$ case with $N_C=2k$. Here, the mass matrix diagonalizes into $\textbf{k}\oplus\textbf{2}\oplus\cdots\oplus\textbf{2}$, again with all $2\times2$ blocks being anti-diagonal. The $k\times k$ sub-block has non-vanishing determinant
\begin{equation}
    \det \langle W^{\prime\prime}|_{k\times k}\rangle=\frac{N_C+2}{2B_0^2H^{N_C-6}M^4},
\end{equation}
\noindent
where $B_0,H,M$ here refer to the non-zero entries of the gauge invariants for the vacuum ~\cref{eq:NF2evenansatz2} so that indeed there are no massless fermions in the spectrum. The complete list of masses of the fermions, scalars and pseudoscalars are shown in \cref{tab:masses}. The 't Hooft anomaly coefficients vanish in the UV, as expected given there are no massless fermions in the IR. Explicitly,
\begin{align}
    &\Tr[\U(1)_6] = 0 + \Nc 2 (\Nc - 3) + 2\Nc (3 - \Nc) = 0 ,\nonumber \\
    &\Tr[\U(1)_6^3] = 0 + \Nc 2 (\Nc - 3)^3 + 2\Nc (3 - \Nc)^3 = 0 ,\nonumber \\
    &\Tr[\U(1)_6 \, \Sp(\Nc - 4)^2] = 0 ,\nonumber \\
    &\Tr[\U(1)_6 \, \SU(2)_D^2] = \frac{1}{2}(2 \Nc (\Nc - 3) + 2 \Nc (3 - \Nc))\nonumber\\
    &\hspace{2.65cm}= 0.
\end{align}
The second to last trace vanishes trivially in analogy to the $\Nf = 1$ case since the $\Sp(\Nc - 4)$ subgroup only effects $\widetilde{Q}$ for $i < \Nc - 3$ while the $\U(1)_6$ is only nonzero for $i \geq \Nc - 3$ for the same field.

\section{Non-Supersymmetric Limit}
\label{sec:nonsusy}
We now turn to applying the above results to the non-supersymmetric gauge theories defined by $(\Nc,\Nf)$ as discussed in~\cref{sec:intro}. As discussed in that same section, this is done by considering the $m\rightarrow\infty$ limit in the vacua above. By assuming that no phase transition occurs as $m$ and $\Lambda$ swap hierarchies, we can conjecture several statements about the non-SUSY theories. In particular, we conjecture the surviving global symmetries and massless fermion content of the non-SUSY theories are identical to those of the SUSY theories above when deformed by AMSB. We summarize these results in the current work together with those of~\cite{Csaki:2021xhi} in Table~\ref{tab:symmetry}.

To bolster the conjecture, we can analyze the non-SUSY theories to determine massless fermion candidates. For $\SU(2k+1)$, we need massless composite fermions to match the anomalies under the unbroken symmetries. For $N_F=0$, supersymmetry is dynamically broken. The massless fermions are $(A\widetilde{Q})_i \widetilde{Q}_j \propto J_{ij}$ for $i=1,\cdots,2k-4$, as well as a singlet $(A\widetilde{Q})_{2k-3} \widetilde{Q}_{2k-3}$ \cite{Csaki:2021xhi}. Here, parentheses indicate the contraction of spinor indices.
\begin{table*}
\def\arraystretch{1.7}
\begin{tabular}{| c | c | c | c |}
    \hline
    $\Nc$ & $G_{\rm global}$  & $H_{\rm global}$  & massless fermions \\ \hline\hline
    \multicolumn{4}{|c|}{$\Nf=0$} \\ \hline
    Odd & $\SU(\Nc-4)\times \U(1)$ & $\Sp(\Nc-5)\times \U(1)'$ 
    & {$\scalebox{0.5}{\yng(1)}+ 1$} \\ \hline
    Even & $\SU(\Nc-4)\times \U(1)$ & $\Sp(\Nc-4)$ 
    & none \\ \hline\hline
    \multicolumn{4}{|c|}{$\Nf=1$} \\ \hline
    Odd & $\SU(\Nc-3)\times \U(1)_1 \times \U(1)_2$ & $\Sp(\Nc-3)\times \U(1)_2$ 
    & {$\scalebox{0.5}{\yng(1)}$} \\ \hline
    Even & $\SU(\Nc-3)\times \U(1)_1 \times \U(1)_2$ & $\Sp(\Nc-4)\times \U(1)_3$ 
    & none \\ \hline\hline
    \multicolumn{4}{|c|}{$\Nf=2$} \\ \hline
    Odd & $\SU(\Nc-2)\times \SU(2) \times \U(1)_1 \times \U(1)_2$ & $\Sp(\Nc-3)\times \U(1)_4 \times \U(1)_5$ 
    & {$\scalebox{0.5}{\yng(1)}$} \\ \hline
    Even & $\SU(\Nc-2)\times \SU(2) \times \U(1)_1 \times \U(1)_2$ & $\Sp(\Nc-4)\times \SU(2) \times \U(1)_6$ 
    & none \\ \hline
\end{tabular}
\caption{Conjectured global symmetries and massless fermion content of the non-supersymmetric gauge theories. Representations of massless fermions are shown in Young tableaux of unbroken $\Sp$ groups.}
\label{tab:symmetry}
\end{table*}

On the other hand for $N_F = 1, 2$ studied here, anomalies are matched by fermion components of mesons (mesinos) $M_i=(Q\widetilde{Q}_i)$ $(i=1,\cdots, 2k-2)$ for $N_F=1$ or $M_i^2=(Q^2\widetilde{Q}_i)$ $(i=2, \cdots, 2k-1)$ for $N_F=2$. At the first sight, it does not appear that such a spectrum can remain in the non-SUSY limit $m\rightarrow \infty$ because the mesinos are made of a quark and a squark. However, they can be replaced by $(A \widetilde{Q}_i)^* Q$ or $(A \widetilde{Q}_i)^* Q^2$ which consist only of fermions and have the same quantum numbers as the mesinos under the unbroken symmetries. This is because $H_{ij} = A \widetilde{Q}_i \widetilde{Q}_j\neq 0$ breaks $SU(2k-3+N_F)$ to $Sp(2k-2)$, and hence $(A \widetilde{Q}_i)^*$ has the same quantum number as $\widetilde{Q}_i$. We find it highly non-trivial that we do find candidates for massless composite fermions in the non-SUSY limit, giving credence to the idea that both limits are smoothly connected without a phase transition.

It is interesting to note that the symmetry breaking patterns in Table~\ref{tab:symmetry} are not those expected in the tumbling hypothesis. The idea behind tumbling is to identify fermion bilinear condensates that break the gauge group based on the Most Attractive Channel (MAC). The MAC is based on a simple one-gluon exchange potential with the coefficient
\begin{align}
    [(T^a)_{R_1}(T^a)_{R_2}]_{R_3}
    = \frac{1}{2} [C_2(R_3) - C_2(R_1) - C_2(R_2)],
\end{align}
where $R_1$ and $R_2$ are the representations of the fermions while $R_3$ is the representation of the bilinear operator called the ``channel'' and $C_2$ is the quadratic Casimir for a given representation. The  attractive channels are 
\begin{align}
    [\widetilde{Q}Q]_1 &: 0 - C_2({\scalebox{0.5}{\yng(1)}}) - C_2(\overline{\scalebox{0.5}{\yng(1)}}) 
    = -\frac{(\Nc-1)(\Nc+1)}{\Nc},
\end{align}
\begin{align}
    [A \widetilde{Q}]_{\scalebox{0.4}{\,\yng(1)}} &: C_2({\scalebox{0.5}{\yng(1)}}) - C_2({\scalebox{0.5}{\yng(1)}}) - C_2(\overline{\scalebox{0.5}{\yng(1)}}) 
    = -\frac{(\Nc-2)(\Nc+1)}{\Nc},
\end{align}
and
\begin{align}
    [Q Q]_{{\scalebox{0.4}{ \yng(1,1)}}} &: C_2\left(\hspace{0.1 em} {\tiny\Yvcentermath1 \yng(1,1)}\hspace{0.1 em}\right) - C_2({\scalebox{0.5}{\yng(1)}}) - C_2({\scalebox{0.5}{\yng(1)}}) = -\frac{\Nc+1}{\Nc}.
\end{align}
Based on this, the tumbling hypothesis \cite{Georgi:1979md,Raby:1979my,Goity:1985tf} would predict that $\Nf (\widetilde{Q}_i Q^j) \propto \delta_i^j$ would condense first, breaking $\SU(\Nc+\Nf-4) \times \SU(\Nf) \times \U(1)^2 \rightarrow \SU(\Nc-4) \times \SU(\Nf) \times \U(1)^2$. Then $(A_{\alpha\beta}\widetilde{Q}_i^\beta) \propto \delta_i^\alpha$ would condense next, breaking the $\SU(\Nc)$ gauge group to $\SU(4)$ while preserving the diagonal $\SU(\Nc-4)$ subgroup of the gauge and flavor groups. The symmetric components of $\widetilde{Q}$ remain massless. The remaining fermions are $A \supset (4, \overline{\Nc-4}) \oplus (6, 1)$ and $\widetilde{Q} \supset (\bar{4}, \Nc-4)$ and hence are vector-like. The unbroken symmetry is $\SU(N_C-4)\times SU(N_F) \times \U(1)^2$. Massless fermions can be identified as composite fermions $(A Q^{\{i,})Q^{j\}}$ in the symmetric representation of $\SU(\Nc-4)$. 

Our results with SUSY$+$AMSB suggest competitions among various attractive channels. In all cases studied, the $[Q Q]_{{\scalebox{0.4}{ \yng(1,1)}}}$ channel condenses. However, the $[Q \widetilde{Q}]_1$ channel appears to be in competition with it. 
Focusing on the non-Abelian subgroups, for $\SU(2k+1)$, the $[Q Q]_{{\scalebox{0.4}{ \yng(1,1)}}}$ channel appears to ``win'' and the condensates prefer the largest possible symplectic symmetry. In the case with $N_F=1$, $\langle \widetilde{Q}_i Q^j\rangle \propto J_{ij}$ breaks the $\SU(2k-2)$ to $\Sp(2k-2)$ without a condensate in the $[\widetilde{Q}_i Q^j]_1$ channel. For $N_F=2$, the same channel breaks $\SU(2k-1)\times \SU(2)$ to $\Sp(2k-2)$ which allows for a condensate in the $[\widetilde{Q}_i Q^j]_1$ channel but only for one of the flavors. On the other hand for $\SU(2k)$, the $[\widetilde{Q}_i Q^j]_1$ channel ``wins'' and breaks $\SU(2k+N_F-4)$ to $\SU(2k-4)$ which is further broken to $\Sp(2k-4)$ by condensates in the $[Q Q]_{{\scalebox{0.4}{ \yng(1,1)}}}$ channel. The $[A \widetilde{Q}]_{\scalebox{0.4}{\,\yng(1)}}$ channel appears to have condensates in all cases. 

Note that order parameters of symmetry breaking persist in the form of fermion bilinear or quadrilinear condensates in the non-SUSY limits. The fermion bilinear $\psi_Q\psi_{\tilde{Q}}$ is the $F$-component of the meson chiral superfield $Q\tilde{Q}\ni \theta^2 \psi_Q\psi_{\tilde{Q}}$ in the UV description. In the IR, it can be worked out as $Q\tilde{Q}\ni \theta^2 (Q F_{\tilde{Q}} + F_Q \tilde{Q})$ \cite{KMN}. Let us take $N_C=2k$ and $N_F=1$ as an example, and let us focus on dependencies on $m$ and $\Lambda$, while dropping all coefficients. The overall expectation value of the fields scales with $\rho \propto m^{-1/2k} \Lambda^{1+1/2k}$. In addition, they also acquire an $F$-component as $F \propto \partial W/\partial \rho \propto m^{1-1/2k} \Lambda^{1+1/2k} \propto m \rho$. The meson field, made of scalar quarks, behaves as $M_{2k-3} = Q \tilde{Q}_{2k-3} \propto m^{-1/k} \Lambda^{2+1/k}$ and damps as $m \rightarrow \infty$. On the other hand, the fermion bi-linear condensate takes the form $\psi_Q \psi_{\tilde{Q}_{2k-3}} \propto m^{1-1/k} \Lambda^{2+1/k}$ and increases in the non-SUSY limit. In order to compare to the expected behavior in the non-SUSY limit, note that $\Lambda_{\cancel{SUSY}}$ can be obtained by matching the one-loop beta functions with $b_0^{SUSY} = 6k$ and $b_0^{\cancel{SUSY}} = \frac{22}{3}k$, $\Lambda = m^{-2/9} \Lambda_{\cancel{SUSY}}^{11/9}$. Therefore, the fermion bilinear condensate behaves as $\psi_Q \psi_{\tilde{Q}_{2k-3}} \propto m^{\frac{5}{9}-\frac{11}{9k}} \Lambda_{\cancel{SUSY}}^{\frac{22}{9}+\frac{11}{9k}}$ when $m \ll \Lambda_{\cancel{SUSY}}$. It is easy to imagine that it would smoothly connect to $\psi_Q \psi_{\tilde{Q}_{2k-3}} \propto \Lambda_{\cancel{SUSY}}^3$ when $m \gg \Lambda_{\cancel{SUSY}}$. In addition, we expect another order parameter $({\bar{\psi}_{\tilde{Q}}\bar{\psi}_{\tilde{Q}}})_{[\alpha,\beta]}^{[i,j]} ({\psi_{\tilde{Q}}}{\psi_{\tilde{Q}}})^{[\alpha,\beta]}_{[k,l]} \propto J^{ij} J_{kl}$. Indeed, the $D$-component of the superfield ${\tilde{Q}}^{\dagger[i,}_{[\alpha,} {\tilde{Q}}^{\dagger j]}_{\beta]} {\tilde{Q}}^{[\alpha,}_{[k,} {\tilde{Q}}^{\beta]}_{l]}$ fulfills this role. The scalar quadrilinear damps as $\rho^4 \propto m^{-2/k} \Lambda^{4+2/k}$ while the fermion quadrilinear increases as $\rho^4 \propto m^{2-2/k} \Lambda^{4+2/k}$. Combining the bilinear and quadrilinear condensates of fermions, the symmetries are broken in the same way as in the small SUSY breaking $m\ll \Lambda$. This way, we can expect a smooth cross-over from the near-SUSY to the non-SUSY limit.

We hope future developments in lattice gauge theories will allow for studies of these theories and show which symmetry breaking patterns prove true in the non-SUSY limits.

\section{Numerics}
\label{sec:num}

\begin{figure}[]
    \centering
    \includegraphics[clip,scale=0.75]{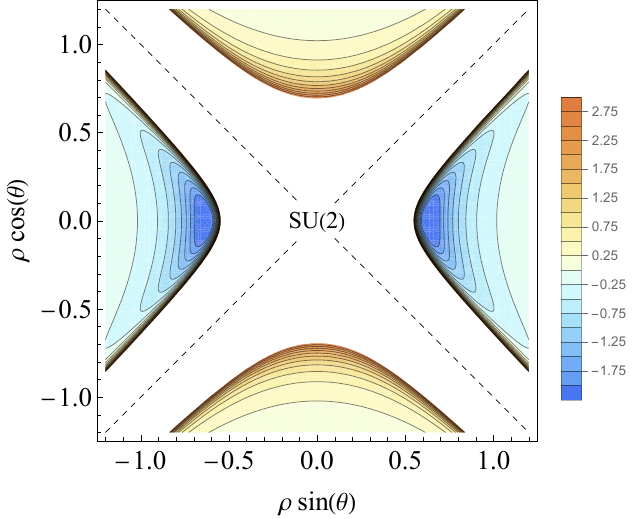}
    \caption{A plot of the $N_f=2$ scalar potential for $N_C=4$ over two parameters $\rho,\theta$ and $\Lambda=1,m=1/2$. The dashed lines indicate the regions where the theory exhibits an enhanced gauge $\text{SU}(2)$ symmetry. The parameter $\theta$ interpolates the minimum configuration ~\cref{eq:NF2evenansatz_uv} through the region of enhanced symmetry when $\theta$ is some integral multiple of $\pi/4$.}
    \label{fig:nf2_potential}
\end{figure}
In this section, we conduct numerical checks to verify our vacuum ansatz is valid for $N_C< 9$ for all gauge theories presented above. Specifically, we minimize the scalar potential ~\cref{eq:AMSBpot} of such theories and confirm that the ansatz for the minimum, mass spectra, and therefore symmetry breaking patterns properly match. 

A thorough analysis of these theories for any $N_C$ is not feasible with our current analytical and numerical methods. The complexity of the gauge invariants and exponential increase in parameters limits the study of higher $N_C$ theories with the techniques described here.

Throughout the numerical analysis, we check for the possibility of multiple vacua in each theory. Due to the complicated structure of the scalar potential, the existence of a single unique minimum is not guaranteed. The strength of using numerics is that we can search for candidate minima in a more robust way. Since theories with a large number of parameters typically cannot be minimized analytically and sometimes even a full D-flat parametrization is not tractable, these numerics provide a method by which to systematically search for possible vacua.

To do these numerical calculations, we use the  library \texttt{SymEngine} through its Python wrapper as this provides symbolic and numerical tools for manipulating algebraic expressions. We symbolically compute our gauge invariants and construct our scalar potential. Numerically, \texttt{SymEngine} evaluates expressions to any desired floating-point precision. At first, this increases confidence that our results match with low numerical error. However, for the $N_F=2$ case with even $N_C$, numerical stability of the superpotential requires higher than the standard \texttt{float64} precision, which we attribute towards sensitive cancellation of the two terms that appear in the denominator of ~\cref{eq:Nf2super}.

After obtaining a symbolic form of the scalar potential in \texttt{SymEngine}, we substitute field parametrizations for which we would like to search for a potential vacuum. We then numerically search for local minima using the Broyden-Fletcher-Goldfarb-Shanno (BFGS) algorithm. As a second-order minimization algorithm, each iteration towards a minimum is determined by both gradient and Hessian estimates. Including the Hessian improves the search due to the geometrical nature of $\text{SU}(N_C)$ and the moduli space structure produced from the matter representations. Moreover, since BFGS is an \textit{unconstrained} method, providing a parametrization of D-flat configurations greatly improves the convergence of BFGS to a local minimum. 

In cases where we cannot find a good parametrization of the D-flat conditions, we can add a Lagrange multiplier $\lambda$ to the potential so that we should minimize the function
\begin{align}
    f_1(A,Q,\widetilde{Q},\lambda)=V_F+V_{AMSB}+\lambda V_D,
\end{align}
\noindent
which is a slightly adjusted form of the scalar potential in ~\cref{eq:AMSBpot}. In trying to minimize $f_1$, BFGS can fail to find a minimum. Since the Lagrange multiplier $\lambda$ is linear in $f_1$ in order to enforce the constraint $V_D=0$, then arbitrarily negative $\lambda$ will lead to a tachyonic direction in which $f_1\to-\infty$. We can circumvent this instability by minimizing the function
\begin{align}
    f_2(A,Q,\widetilde{Q},\lambda)=V_F+V_{AMSB}+(a+|\lambda|)V_D,
    \label{eq:lagrange}
\end{align}
\noindent
with $a>0$ a constant that softly imposes a vanishing D-term. Since $f_2$ is no longer linear in $\lambda$, minimizing $f_2$ can prefer a solution in which $\lambda=0$ so that only the term $aV_D$ will push the function towards D-flat directions. We find that this method does not lead to a minimum of the potential $V$ that is guaranteed to be D-flat, since if BFGS sends $\lambda\to0$, then in general we would have to set $a\to\infty$ to obtain a D-flat vacuum solution.

For theories with $\Nf = 1$ as well as those with $N_F = 2$ odd $N_C$, we implement a general field parametrization using only the global symmetries of the theory to rotate away extraneous parameters. We enforce the D-flat condition in our minimization algorithm using the Lagrange multiplier constraint. We test both this parametrization as well as one already constrained to be D-flat and both found vacua equivalent to \cref{eq:NF1oddansatz,eq:NF1evenansatz,eq:NF2oddansatz} up to gauge and global rotations. Similarly, we obtain identical mass spectra to those described in \cref{tab:masses}.

In the case of theories with $N_F=2$ and even $N_C$, providing a general parametrization which is not D-flat using a Lagrange multiplier as in ~\cref{eq:lagrange} did not result in a local minimum of $f_2$ that is D-flat. Instead, we use configurations that were already D-flat so that the BFGS algorithm converged properly. We note that this type of theory can lead to multiple distinct minima since regions of enhanced symmetry split up regions of finite value of the scalar potential, with AMSB producing at most a single minimum in each region. With many checks of different configurations, we find only a single minimum, with other regions not receiving large enough contributions from the AMSB to produce another minimum (see Figure \ref{fig:nf2_potential}).

\vspace{-5mm}
\section{Conclusions}
\label{sec:conc}
In this work we obtained exact results in a family of $\mathcal{N}=1$ supersymmetric chiral gauge theories softly broken by small AMSB, and conjectured they smoothly connect to the non-supersymmetric limits. In particular, we identified unbroken global symmetries that survive in the IR of these theories. The full symmetry breaking patterns are summarized in Table~\ref{tab:symmetry}. Interestingly, these patterns differ from what the tumbling hypothesis suggests. For odd $\Nc$, these symmetries are anomalous and there exist massless fermions in the IR to satisfy 't Hooft anomaly matching. The massless composite fermions are the fermionic components of the meson superfield $\tilde{Q} Q$, which are not expected to remain light, yet they can smoothly turn into the composite fermion $(A \widetilde{Q})^* Q$ in the non-SUSY limit. This point is quite non-trivial and supports the idea that the small SUSY breaking $m\ll \Lambda$ is smoothly connected to the non-SUSY limit $m\rightarrow \infty$ without a phase transition. For even $\Nc$, the symmetries are not anomalous and correspondingly we did not find massless fermions in the IR. 

\vspace{-3mm}
\begin{acknowledgments}
JML was co-funded by the European Union and supported by the Czech Ministry of Education, Youth and Sports (Project No.~FORTE – CZ.02.01.01/00/22\_008/0004632), as well as by the Deutsche Forschungsgemeinschaft under Germany’s Excellence Strategy - EXC 2121 “Quantum Universe” - 390833306. The work of B.A.S. is supported by the NSF GRFP Fellowship and BSF-2018140. 
    This material is based upon work supported by the National Science Foundation Graduate Research Fellowship Program under Grant No.\ DGE 2146752. Any opinions, findings, and conclusions or recommendations expressed in this material are those of the author(s) and do not necessarily reflect the views of the National Science Foundation. The work of HM was supported by the US DOE Contract No. DE-AC02-05CH11231, by the NSF grant PHY-2210390, by the JSPS Grant-in-Aid for Scientific Research JP23K03382, MEXT Grant-in-Aid for Transformative Research Areas (A) JP20H05850, JP20A203, BSF-2018140, by the World Premier International Research Center Initiative, MEXT, Japan, Hamamatsu Photonics, K.K, and Tokyo Dome Corporation.
\end{acknowledgments}

\clearpage

\onecolumngrid
\appendix
\section{Global Charges\label{sec:charges}}

In \cref{tab:charges2}, we present the field charges under the unbroken global U(1) symmetry groups present in the vacua discussed above. Note that some Abelian symmetries and gauge-invariant polynomials are defined only in theories with $\Nc$ odd (or even), leading to cases where a given polynomial is excluded from participating in a particular global symmetry. Such cases are denoted by the null sign $\varnothing$ to distinguish them from fields that do participate in a given symmetry but simply have 0 charge.

\begin{table*}[ht]
\def\arraystretch{2}
\begin{tabular}{| c || c | c | c | c | } 
 \hline
  &  $\U(1)_3$ & $\U(1)_4$ & $\U(1)_5$ & $\U(1)_6$ \\
  \hline
  $\Nc$ & Even & Odd & Odd & Even \\
  \hline
  $\Nf$ & 1 & 2 & 2 & 2  \\ 
  \hline\hline
  $A_{\alpha\beta}$ & 0 & $-\frac{2(\Nc-2)}{\Nc}$ & $-4$ & 0 \\
  \hline
  $Q^a_\alpha$ & $\Nc - 3 $ & $\frac{\Nc^2-2\Nc+2}{\Nc}$ & \begin{tabular}{@{}cc@{}}
    $-2$, & $a = 1$\\
    $2(\Nc-1)$, & $a = 2$
        \end{tabular} & $\Nc - 3 $ \\
    \hline 
    $\widetilde{Q}^\alpha_i$ & \begin{tabular}{@{}cc@{}}
    0, & $i\neq\Nc-3$\\
    $3-\Nc$, & $i=\Nc-3$
       \end{tabular} & \begin{tabular}{@{}cc@{}}
    $\frac{\Nc-2}{\Nc}$, & $i \neq 1$\\
    $-\frac{\Nc^2 -3\Nc+2}{\Nc}$, & $i = 1$
\end{tabular} & 2 & \begin{tabular}{@{}cc@{}}
    0, & $i < \Nc-3$\\
    $3-\Nc$, & $i \geq \Nc-3$
\end{tabular} \\
 \hline
 \hline
 $H_{ij}$ & \begin{tabular}{@{}cc@{}}
    0, & $i,j\neq\Nc-3$\\
    $6-2\Nc$, & $i||j=\Nc-3$
\end{tabular} & \begin{tabular}{@{}cc@{}}
    $0$, & $i,j \neq 1$\\
    $2-\Nc$, & $i||j = 1$
\end{tabular} & 0 & \begin{tabular}{@{}cc@{}}
    0, & $i,j < \Nc-3$\\
    $6-2\Nc$, & $\{i,j\} \in \{\Nc-3, \Nc-2\}$\\
    $3-\Nc$, & otherwise
\end{tabular} \\
  \hline
  $M^a_i$ & \begin{tabular}{@{}cc@{}}
    $\Nc-3$, & $i\neq\Nc-3$\\
    0, & $i=\Nc-3$
\end{tabular} & \begin{tabular}{@{}cc@{}}
    $0$, & $i = 1$\\
    $\Nc-2$, & $i \neq 1$
\end{tabular} & \begin{tabular}{@{}cc@{}}
    $0$, & $a = 1$\\
    $2\Nc$, & $a = 2$
\end{tabular} & \begin{tabular}{@{}cc@{}}
    $\Nc-3$, & $i < \Nc-3$\\
    0, & $i \geq \Nc-3$
\end{tabular} \\
\hline
$B_0$ & 0 & $\varnothing$ & $\varnothing$ & 0 \\
\hline
$B^a_1$ & $\varnothing$ & 0 & \begin{tabular}{@{}cc@{}}
    $-2\Nc$, & $a = 1$\\
    $0$, & $a = 2$
\end{tabular} & $\varnothing$ \\
\hline
$B_2$ & $\varnothing$ & $\varnothing$ & $\varnothing$ & 2 \\
 \hline
\end{tabular}
\caption{Charges of the gauge-invariant polynomials for the unbroken global Abelian symmetries. The $\varnothing$ entries indicate that the global symmetry is not relevant for the field in question. }\label{tab:charges2}
\end{table*}

\section{Fermion and Scalar Masses\label{sec:masses}}

\cref{tab:masses} lists all of the masses of the fermion, scalar, and pseudoscalars along the D-flat directions for each of the four classes of theories. In particular, we do not include particles corresponding to massive gauge bosons, gauginos, or D-scalars. We expect these masses to be very heavy compared to the light spectra, e.g. $m_{G, \lambda, D} \sim \mathcal{O}(g\cdot\rho)$ where g is the gauge coupling. The first column lists the mass of the Higgs. The second column lists the masses of the Nambu-Goldstone supermultiplets of the spontaneously broken global symmetries, while the third lists the massless modes. The last row in each section enumerates the representations of the spectra under the unbroken global non-abelian subgroup of $\mathcal{G}_{UV}$ along the vev. $N_F=2$, $N_C$ even acts as a special case in which two extra massive singlets are present in the spectrum. We note that for the odd $N_C$ cases, the theory predicts an increasing number of massless fermions for increasing $N_C$. Cases with even $N_C$ do not have any massless fermions. 

The supertrace condition requires that 
\begin{equation}
    \text{Str}(M^2) = -2 \,\text{tr} (M_f^2) + \text{tr} (M_b^2) = 0.
\end{equation}
This is fulfilled for each column of masses shown on this table.

The $\varnothing$ symbol denotes that there are no fields in a particular category.

\begin{table*}[ht]
    \centering
    \def\arraystretch{1.7}
    \begin{tabular}{|c|c|c|c|c|}
        \hline
         &  Higgs  & $m_{(NG)}$  & $m_f = 0$ & $m_s$ \\
        \hline\hline
        \multicolumn{5}{|c|}{$\Nc$ Odd, $\Nf = 1$} \\
        \hline
        $M_{b, \,\rm heavy}^2$ & $\frac{\Nc}{\Nc - 1}(\Nc + 1)^2m^2$ & $2\frac{(\Nc + 1)^2}{(\Nc - 1)^2}m^2$ & 0 & $\varnothing$ \\
        \hline
        $M_f^2$ & $(\Nc + 1)^2m^2$ & $\frac{(\Nc + 1)^2}{(\Nc - 1)^2}m^2$ & 0 & $\varnothing$ \\
        \hline
        $M_{b, \, \rm light}^2$ & $\frac{\Nc - 2}{\Nc - 1}(\Nc + 1)^2m^2$ & $0$ & $0$ & $\varnothing$ \\
        \hline
        $\Sp(N_c-3)$ & 1 & ${\tiny\Yvcentermath1\yng(1,1)}+1$ & ${\tiny\Yvcentermath1 \yng(1)}$ & $\varnothing$ \\
        \hline\hline 
        \multicolumn{5}{|c|}{$\Nc$ Even, $\Nf = 1$} \\
        \hline
        $M_{b, \, \rm heavy}^2$ & $\frac{\Nc}{\Nc - 1}(\Nc + 1)^2m^2$ & $2\frac{(\Nc + 1)^2}{(\Nc - 1)^2}m^2$ & $\varnothing$ & $\varnothing$ \\
        \hline
        $M_f^2$ & $(\Nc + 1)^2m^2$ & $\frac{(\Nc + 1)^2}{(\Nc - 1)^2}m^2$ & $\varnothing$ & $\varnothing$ \\
        \hline
        $M_{b, \, \rm light}^2$ & $\frac{\Nc-2}{\Nc - 1}(\Nc + 1)^2m^2$ & 0 & $\varnothing$ & $\varnothing$ \\
        \hline
        $\Sp(N_c-4)$ & 1 & ${\tiny\Yvcentermath1 \yng(1,1)}+2\ {\tiny\Yvcentermath1 \yng(1)}+2\times 1$ & $\varnothing$ & $\varnothing$ \\
        \hline\hline
        \multicolumn{5}{|c|}{$\Nc$ Odd, $\Nf = 2$} \\
        \hline
        $M_{b, \, \rm heavy}^2$ & $\frac{2\Nc}{2\Nc - 1}(2\Nc + 1)^2m^2$ & $2\frac{(2\Nc + 1)^2}{(2\Nc - 1)^2}m^2$ & 0 & $\varnothing$ \\
        \hline
        $M_f^2$ & $(2\Nc + 1)^2m^2$ & $\frac{(2N_C+1)^2}{(2N_C-1)^2}m^2$ & 0 & $\varnothing$ \\
        \hline
        $M_{b, \, \rm light}^2$ & $\frac{2N_C-2}{2\Nc - 1}(2\Nc + 1)^2m^2$ & 0 & 0 & $\varnothing$ \\
        \hline
        $\Sp(N_c-3)$ & 1 & ${\tiny\Yvcentermath1 \yng(1,1)}+2\ {\tiny\Yvcentermath1 \yng(1)}+4\times 1$ & ${\tiny\Yvcentermath1 \yng(1)}$ & $\varnothing$ \\
        \hline\hline 
        \multicolumn{5}{|c|}{$\Nc$ Even, $\Nf = 2$} \\
        \hline
        $M_{b, \, \rm heavy}^2$ & $\frac{2\Nc}{2\Nc - 1}(2\Nc + 1)^2m^2$ & 2$\frac{(2\Nc + 1)^2}{(2\Nc - 1)^2}m^2$ & $\varnothing$ & 6$\frac{(2\Nc + 1)^2}{(2\Nc - 1)^2}m^2$ \\
        \hline
        $M_f^2$ & $(2\Nc + 1)^2m^2$ & $\frac{(2\Nc + 1)^2}{(2\Nc - 1)^2}m^2$ & $\varnothing$ & $4\frac{(2\Nc+1)^2}{(2\Nc-1)^2}m^2$ \\
        \hline
        $M_{b, \, \rm light}^2$ & $\frac{2\Nc - 2}{2\Nc - 1}(2\Nc + 1)^2m^2$ & 0 & $\varnothing$ & $2\frac{(2\Nc + 1)^2}{(2\Nc - 1)^2}m^2$ \\
        \hline
        $\Sp(N_c-4)\times \SU(2)_D$ & 1 & $( \, {\tiny\Yvcentermath1 \yng(1,1)},1)+2({\tiny\Yvcentermath1 \yng(1)},{\tiny\Yvcentermath1 \yng(1)})+(1,{\tiny\Yvcentermath1 \yng(2)})+2\times 1$ & $\varnothing$ & $2\times 1$ \\
        \hline
    \end{tabular}
    \caption{This table lists all the masses of $D$-flat directions and their representations under the unbroken global symmetries at the tree level for each of the four cases. We have divided up the masses of the scalars to those heavier than the fermions and lighter than the fermions, denoted by $M_{b, \, \rm heavy}^2$ and $M_{b, \, \rm light}^2$. Higgs refers to the supermultiplet in the field direction of the vev. NG refers to Nambu--Goldstone supermultiplets for spontaneously broken global symmetries. Only the odd $\Nc$ cases have massless fermions; their scalar partners acquire positive mass squared from two-loop AMSB. $\U(1)$ quantum numbers are omitted for brevity. The $\varnothing$ entries denote that there are no masses of that type in that case. All cases shown here fulfill the supertrace condition for each column. }
    \label{tab:masses}
\end{table*}
\twocolumngrid

\bibliography{AMSB_Refs}

\end{document}